\documentclass[aps,prb,twocolumn,superscriptaddress,noeprint,nofootinbib,longbibliography,floatfix, nobalancelastpage]{revtex4-2}

\usepackage{graphicx}%
\usepackage{bm}%
\usepackage{amsmath, amssymb, mathtools}
\usepackage{booktabs}
\usepackage{mathrsfs}
\usepackage[normalem]{ulem}
\usepackage{pifont}
\usepackage[mathscr]{euscript}
\usepackage{hyperref}
\usepackage{soul}
\usepackage{dsfont}
\usepackage[inline]{enumitem} %
\usepackage[dvipsnames]{xcolor}
\usepackage{tikz}

\usepackage{algpseudocode}
\newenvironment{algorithm*}[1][!t]{%
  \begin{figure*}[#1]\centering\begin{minipage}{0.95\textwidth}%
  \hrule height 0.8pt\vspace{6pt}%
}{%
  \vspace{6pt}\hrule height 0.8pt%
  \end{minipage}\end{figure*}%
}
\newcommand{\AlgCaption}[1]{\noindent\textbf{Algorithm}\quad#1\par\vspace{2pt}\hrule\vspace{4pt}}

\definecolor{DarkerGray}{HTML}{a1a1a1}
\definecolor{LiteGray}{HTML}{e7e7e7}

\tikzset{matter spin/.style={circle,thick,draw,DarkerGray,line width=1,fill=LiteGray,scale=0.75}}

\hypersetup{
    colorlinks = true,
    linkcolor  = blue,
    citecolor  = blue,
    urlcolor   = blue,
    linktocpage=true
}

\usepackage{physics}
\usepackage{comment}

\def\sz{\sigma^{\rm z}}
\def\sx{\sigma^{\rm x}}
\def\sy{\sigma^{\rm y}}

\begin{document}

\title{Autonomous phase discovery}

\author{Shiyu Zhou}
\affiliation{Perimeter Institute for Theoretical Physics, Waterloo, Ontario N2L 2Y5, Canada}
\affiliation{Department of Physics, Harvard University, Cambridge, Massachusetts 02138, USA}

\author{Yuxuan Zhang}
\affiliation{Institute of Physics, Ecole Polytechnique Federale de Lausanne, Lausanne, Switzerland}
\affiliation{Department of Physics, Princeton University, Princeton, New Jersey, USA}

\author{Sebastian Wetzel}
\affiliation{Perimeter Institute for Theoretical Physics, Waterloo, Ontario N2L 2Y5, Canada}
\affiliation{Department of Physics \& Astronomy, University of Waterloo, Ontario, N2L 3G1, Canada}

\author{Roger Melko}
\affiliation{Perimeter Institute for Theoretical Physics, Waterloo, Ontario N2L 2Y5, Canada}
\affiliation{Department of Physics \& Astronomy, University of Waterloo, Ontario, N2L 3G1, Canada}

\author{Xiu-Zhe Luo}
\email{paper@rogerluo.dev}
\affiliation{Perimeter Institute for Theoretical Physics, Waterloo, Ontario N2L 2Y5, Canada}
\affiliation{Department of Physics \& Astronomy, University of Waterloo, Ontario, N2L 3G1, Canada}

\begin{abstract}
Understanding quantum phases of matter has long relied on physicists' intuition and mathematical tools such as symmetry and topology. %
Remarkably successful as these approaches have been, they provide no universal way to explore a Hamiltonian space whose organizing principle is not known in advance.
In this work, we introduce a fully autonomous system combining differentiable programming and unsupervised learning for quantum phase discovery. The search evaluates ground-state data along an adaptive trajectory rather than on a predetermined parameter grid. We demonstrate the system with three different solvers and benchmark it against random sampling at equal ground-state-evaluation budgets. On a generalized cluster chain hosting up to $200$ distinct phases, the search finds up to $25$ more phases at the same budget, and matches random sampling given thirty times its budget. On a $50$-parameter Chern insulator, it reaches sectors not obtained by the simple harmonic constructions considered here, in a family whose inverse problem remains open, while recovering all sectors found by sampling.
Our results establish autonomous, gradient-driven exploration of Hamiltonian space as a practical route to discovering quantum phases without phase labels or a prescribed target phase.
\end{abstract}

\maketitle
\section{Introduction}
\label{sec:intro}
The study of phases of matter underpins broad areas of science and engineering. Understanding exotic quantum matter, such as high-temperature superconductors~\cite{ Bednorz:1986tc,lee2004dopingmottinsulatorphysics}, fractional quantum Hall states~\cite{PhysRevLett.48.1559,Laughlin83}, and, more recently, mixed-state phases%
~\cite{deGroot2022OpenSPT,MaWang2023AverageSPT,Sang2024MixedStatePhases,Wang2025IntrinsicMixedState,Sohal2025NoisyMixedState,Zhang2026MixedState,Su2026MixedStateOrder}, is a central challenge in condensed matter physics, with broad implications for quantum computation and materials science. Conventionally, the understanding of new quantum phases has relied heavily on the insights of physicists using mathematical tools, like symmetry and topology. These approaches have been remarkably successful~\cite{sachdev2011, Wen_2017}, but they offer no universal recipe for discovering new phases: one must search in a continuous Hamiltonian parameter space while recognizing when a point in it realizes physics distinct from what is already known. Programmable quantum simulators sharpen the difficulty as they now realize quantum many-body systems~\cite{Semeghini2021SpinLiquid,satzinger2021realizing,kornjavca2023trimer,evered2025probing,Zhang2022ThermalStates,Anand2023Holographic,Zhang2025Supersonic} richer than pure insight-driven methods can characterize. These challenges have motivated machine-learning approaches that seek to reduce the phase-specific knowledge and manual exploration required.

\begin{figure}[!t]
    \centering
    \includegraphics[width=1\columnwidth]{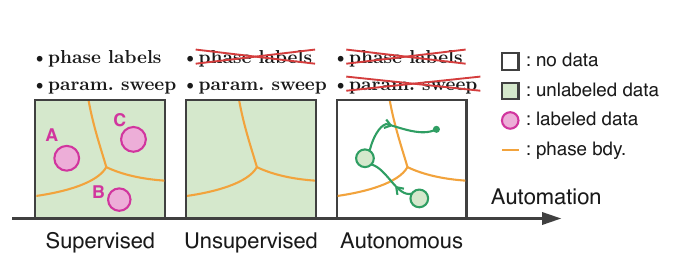}
    \caption{
     Three machine-learning paradigms for mapping quantum phase diagrams with increasing levels of automation. Supervised methods require both phase labels and data from a predetermined sweep of Hamiltonian parameters, whereas unsupervised methods remove the need for phase labels but still rely on such a sweep. In contrast, our autonomous framework requires no prior knowledge of the phases and initializes from data sampled locally around an arbitrary point in Hamiltonian parameter space. The algorithm actively guides the search for new and unknown phases performing measurements or simulations only on the discovery trajectory.
    }
    \label{fig:staircase}
\end{figure}

\begin{figure*}[!t]
\includegraphics[width=0.9\textwidth]{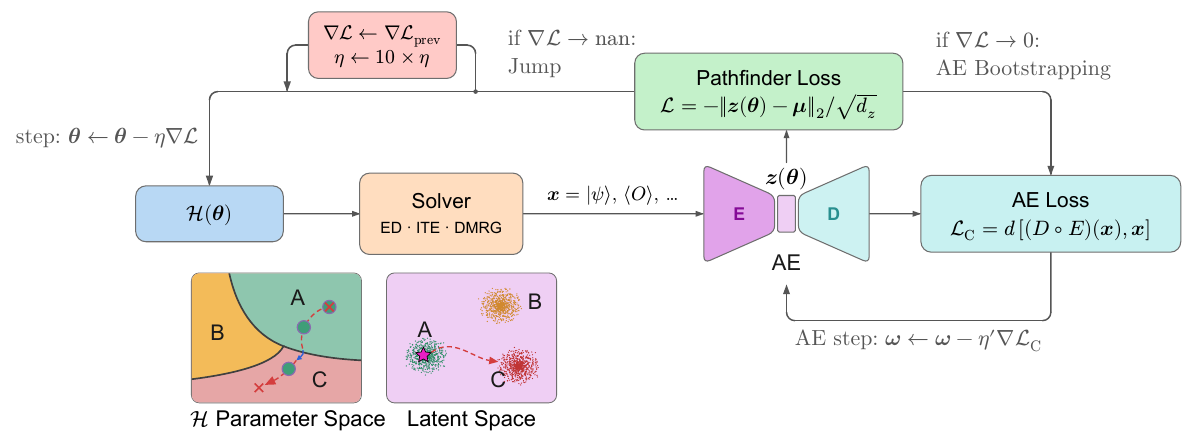}
  \caption{Autonomous phase discovery system. A differentiable Hamiltonian solver (ED, ITE or DMRG here; a closed-form free-fermion backend in Sec.~\ref{sec:scale}) produces ground-state data at the current parameters $\boldsymbol{\theta}$, which the encoder (E) of the autoencoder (AE) compresses into a latent representation $\boldsymbol{z}(\boldsymbol{\theta})$. The pathfinder loss $\mathcal{L} = -\lVert \boldsymbol{z}(\boldsymbol{\theta}) - \boldsymbol{\mu} \rVert_2 / \sqrt{d_z}$, the negative $\ell_2$ distance to the centroid $\boldsymbol{\mu}$ of the visited-phase cluster, is minimized by gradient descent on the Hamiltonian parameters. A diverging gradient, $\nabla \mathcal{L} \rightarrow \mathrm{nan}$, signals a gap closing at a phase boundary and triggers a jump with a boosted learning rate ($\eta \leftarrow 10 \times \eta$). On the other hand, a vanishing gradient, $\nabla \mathcal{L} \rightarrow 0$, triggers the AE bootstrap, retraining the AE on data that include the newly explored region. Bottom left: search trajectory in the Hamiltonian parameter space, from the starting point (cross in phase A) to the discovered phase (cross in phase C). Green markers denote the AE bootstrapping, and the solid blue arc marks the jump across the phase boundary. The same search is shown in the latent space, where ground states of distinct phases (A, B, C) form clusters. The star marks the centroid $\boldsymbol{\mu}$ of the current latent space embedding.}
  \label{fig:schematics}
\end{figure*}

Machine-learning approaches to the study of phases of matter form a natural hierarchy of increasing automation, as shown in~Fig.~\ref{fig:staircase}. Earlier works have demonstrated the effectiveness of supervised learning methods given labeled data~\cite{Carrasquilla_2017, Tanaka2017, PhysRevX.7.031038, vanNieuwenburg_2017, Broecker_2017, PhysRevLett.118.216401, PhysRevB.96.195145, Wetzel2017b, Greitemann2019, Rem_2019, Dawid2020, Miles2023, Arnold2024}, though these require the phases to be known in advance. For certain Hamiltonians with few model parameters, their entire phase diagrams can also be mapped using unsupervised methods without phase labels, such as principal component analysis, diffusion maps, contrastive learning, variational quantum circuits or autoencoders~\cite{Wang_2016, PhysRevE.95.062122, Rodriguez_Nieva_2019, Arnold2021, Greplova2020, Patel2022,Wetzel_2017,Chng2018,Kottmann_2020, Kaming2021,Yin2021,deSchoulepnikoff2025,https://doi.org/10.48550/arxiv.2606.08694,cong2019quantum}.
In approaches based on predetermined parameter sweeps, the region to examine is fixed in advance. For a grid with fixed resolution along each parameter axis, the number of sampling points grows exponentially with the number of parameters. Although classification and generation are distinct learning objectives, they can encode closely related information about the underlying data distribution. In particular, a sufficiently expressive discriminative model can capture the features that distinguish different phases, potentially providing a representation rich enough to probe the structure of phases beyond those encountered during training. This motivates using trained classification models as a tool for discovering previously unidentified phases of matter.%

Meanwhile, the autonomous closed-loop discovery has begun emerging across various fields including materials science and chemistry~\cite{Kusne_2020, Abolhasani_2023, Dai2020}. The underlying steering algorithms are based on reinforcement learning, Bayesian optimization, and active learning to select which experiments or simulations to run next based on historical and real-time data.
In the quantum setting, gradient-based optimization through a differentiable solver has recently been used to search the Hamiltonian parameter space for a prespecified target phase~\cite{https://doi.org/10.48550/arxiv.2509.10438}. As such, a truly autonomous discovery system for searching new and unknown quantum phases that does not rely on any prior knowledge nor an extensive parameter sweeping, has yet to be established.

In this work, we propose a gradient-based phase discovery system that combines differentiable programming with unsupervised learning for phase detection in a closed feedback loop.
It relies on modern differentiable programming~\cite{jax2018github, enzyme2020neurips}, in which ground-state solvers can be written in fully back-propagable forms~\cite{Liao_2019, Xie_2020, Wang2019BackpropagationFriendlyE, TamayoMendoza_2018,Schfer2020}, so that gradients of the loss function can propagate through the solver to the Hamiltonian parameters themselves which determine the phase diagram.
Unsupervised phase detection, in turn, requires no prior knowledge of which phases exist or how they are characterized.
In particular, a sufficiently expressive discriminative model can capture the features that distinguish different phases, potentially providing a representation rich enough to probe the structure of phases beyond those encountered during training.
Previous work has shown that different phases form distinct clusters in the latent space of an autoencoder~\cite{Wetzel_2017, Alexandrou_2020}.
We further recognize that this latent space can also serve as a metric space where the distance from the known phase clusters is a differentiable function of the Hamiltonian parameters, whose gradient can guide the search toward states distinct from the training data. Lastly, we design a feedback loop that coherently ties these components together, steering the search through the Hamiltonian parameter space toward unknown phases.
The feedback loop resembles active learning~\cite{Kusne_2020, settles2009active} in spirit, a machine learning paradigm that aims to label data points in regions of high uncertainty. Our system, however, operates without labels, instead suggesting regions of interest for new measurements or simulations.
The result is an autonomous, gradient-based phase discovery system without the requirements of phase labels, predetermined parameter sweeping, or human supervision.

In the following sections, we first introduce our autonomous system in Sec.~\ref{sec:algo} and then evaluate it in two stages. Sec.~\ref{sec:num} demonstrates its application on three interacting models: the transverse-field Ising chain, the $\mathbb Z_2$ lattice gauge theory, and the transverse-field XXZ chain, the last of which hosts five distinct phases, including a gapless Luttinger liquid. Sec.~\ref{sec:scale} then tests whether our autonomous system improves over uniform random sampling at scale, using strictly equal numbers of ground-state evaluations. We benchmark on a cluster chain with up to $200$ phases and on a $50$-parameter Chern insulator whose reachable phases cannot be enumerated in advance.

\section{The Autonomous System}
\label{sec:algo}

Consider a generic quantum Hamiltonian with multiple phases,
\begin{equation}
    H (\boldsymbol{\theta}) = \sum_i \theta_i \, \hat O_i
    \; ,
\label{eq:ham}
\end{equation}
where $\hat O_i$ is the interaction term, a product of spin-$1/2$ Pauli operators $\sx, \sy, \sz$, and $\theta_i$ is the strength of the interaction $\hat O_i$. Given a fixed set of $\hat O_i$'s, the quantum phases are controlled by the coupling parameters $\boldsymbol{\theta}$ and separated by the boundaries in the space of $\boldsymbol{\theta}$~\footnote{In principle, one can relax the requirement that $\hat O_i$'s are fixed, where the type of pauli operators, the number of pauli operators, and the location of the spin can vary. One can further set loose the requirement of spin-$1/2$. Naturally, quantum phases of matter are then defined by both $\{ \hat O_i \}$, $\boldsymbol{\theta}$ and spin numbers, but in this paper we limit the scope to fixed sets of interactions and leave the more general case to future work}. Historically, the characteristics of the phases and phase boundaries are understood through the lens of the underlying symmetry of the Hamiltonian, the topological properties or entanglement entropies of the ground states, etc~\cite{landau1980statistical, wen2004quantum, RevModPhys.80.517}. However, each of these tools is tied to a specific class of phases, and requires {\it a priori} physical intuition to apply~\cite{Grover_2013, Wen_2002, Balents2010}. No single formalism offers a uniform and universal diagnostic that can be applied without phase-specific inputs.

Our autonomous phase discovery algorithm consists of three fully integrated components: a differentiable Hamiltonian solver, an unsupervised phase detection algorithm called autoencoder, and a phase pathfinder loss, as shown in Fig.~\ref{fig:schematics}. The differentiable Hamiltonian solver provides ground-state information to the autoencoder, which projects it into a latent space. The phase pathfinder then uses this latent representation to direct the search toward unknown phases. The differentiable solver allows the Hamiltonian parameters to become fully trainable variables, so that the gradients of the final loss can propagate all the way back to the parameters. The unsupervised phase detector is pretrained on data generated by the Hamiltonian solver, either before the automatic phase discovery loop begins or when retraining is required during the loop. We first discuss the phase detector and then present the full discovery system.

The unsupervised phase detection algorithm is an autoencoder (AE), a commonly used machine learning model for anomaly detection that produces meaningful latent space embeddings. This algorithm has been widely used to calculate phase diagrams across a broad range of physical systems, including classical spin models, quantum many-body systems, and topological phases of matter~\cite{Wetzel_2017,Chng2018,Kottmann_2020, Kaming2021,Yin2021,deSchoulepnikoff2025,https://doi.org/10.48550/arxiv.2606.08694}. An AE is a neural network model composed of an encoder, a latent space, and a decoder. The encoder (E) compresses the input data $\boldsymbol{x}$ into a latent representation, followed by a decoder (D) that decompresses the latent representation to accurately reconstruct the input data, as shown in Fig.~\ref{fig:schematics}. Typically, an AE exhibits a mirror-symmetric architecture, and the loss function used to train the AE is defined by the distance between the inputs and outputs,
\begin{equation}
    \mathcal L_{\text C} = d \left[ (D \circ E) (\boldsymbol{x}), \, \boldsymbol{x} \right]
    \; .
\end{equation}
The training is purely unsupervised, since no labels are required. The choice of reconstruction loss depends on the input representation. To demonstrate versatility, in this work, we consider inputs for AE from three different Hamiltonian solvers: i) the full wavefunctions from exact diagonalization (ED), ii) the full wavefunctions from imaginary time evolution (ITE), iii) the expectation values of a set of observables of the ground states obtained by density matrix renormalization group (DMRG)~\cite{white1992density,schollwock2005density}. %
For the cases where full wavefunctions are available, it is natural to use fidelity $\mathcal L_{\text C} = 1 - | \bra{\phi} \ket{\psi} |^2$ as the AE loss function, where $\ket \psi$ and $\ket{\phi}$ are the inputs and outputs of AE respectively. As for the observable expectation values, we use mean squared error $\mathcal L_{\text C} = \sum_i ( x_i - y_i )^2 / d$, where $\boldsymbol x$ and $\boldsymbol y$ are the inputs and outputs respectively, and $d$ is the dimension of the data.
We train the autoencoder with gradient descent $\boldsymbol{\omega} \leftarrow \boldsymbol{\omega} - \eta' \nabla_{\boldsymbol{\omega}} \mathcal L_{\text C}$ to minimize the AE loss, where $\boldsymbol{\omega}$ is the AE weights and $\eta'$ is the learning rate.

\begin{algorithm*}
\AlgCaption{Autonomous quantum phase discovery}
\begin{algorithmic}[1]
\Require Hamiltonian $\mathcal H(\boldsymbol{\theta})$, starting point $\boldsymbol{\theta}_0$, sampling radius $r_0$, learning rate $\eta$
\State $\mathcal D \gets$ ground-state data solved at parameters sampled within radius $r_0$ of $\boldsymbol{\theta}_0$
\State train AE on $\mathcal D$; \; $\boldsymbol{\mu} \gets$ centroid of the latent representations of $\mathcal D$
\State $\boldsymbol{\theta} \gets \boldsymbol{\theta}_0$
\While{not converged}
    \State solve the ground state at $\boldsymbol{\theta}$; \; $\boldsymbol z(\boldsymbol{\theta}) \gets$ encoder output
    \State $\mathcal L \gets - \| \boldsymbol z(\boldsymbol{\theta}) - \boldsymbol{\mu} \|_2 \, / \sqrt{d_z}$
    \State $\boldsymbol g \gets \nabla_{\boldsymbol{\theta}} \mathcal L$ \Comment{backpropagate through encoder and solver}
    \If{$\boldsymbol g$ diverges} \Comment{gap closing: phase boundary}
        \State revert $\boldsymbol{\theta}$ to the last finite-gradient step
        \State \textbf{jump}: boost $\eta \gets 10 \times \eta$ temporarily and kick $\boldsymbol{\theta}$ along the last update
    \ElsIf{$\boldsymbol g$ vanishes over a sustained window} \Comment{current phase explored}
        \State $\mathcal D \gets \mathcal D \, \cup$ fresh ground-state data sampled around $\boldsymbol{\theta}$
        \State \textbf{bootstrap}: retrain AE on $\mathcal D$ and update $\boldsymbol{\mu}$
    \Else
        \State $\boldsymbol{\theta} \gets \boldsymbol{\theta} - \eta \, \boldsymbol g$
    \EndIf
\EndWhile
\end{algorithmic}
\end{algorithm*}

A typical AE architecture has a bottleneck in the middle. The bottleneck forces AE to discard information that is not useful for reconstruction, and what survives in the compressed latent space is the essential structure of the input data. Inputs with similar underlying structures can lie close together in the learned latent space, while dissimilar inputs can lie farther apart. This property is what makes latent space useful for examining phase diagrams~\cite{Wetzel_2017, Yin2021, Rodriguez_Nieva_2019, Wang_2016}. Specifically, for an AE trained on the ground-state data from a single phase, the latent representations form a cluster. We define the centroid of the cluster in the latent space as,
\begin{equation}
    \boldsymbol{\mu} = \frac{1}{n} \sum_i \boldsymbol z_i
    \; ,
\end{equation}
where the average is taken over different latent representations, and $n$ is the total number of them. Given an input, the encoder produces a latent representation, and its distance from the centroid $\boldsymbol{\mu}$ serves as a measure of novelty relative to the training data.

The phase pathfinder loss operates directly in the latent space. Given Hamiltonian parameters $\boldsymbol{\theta}$, the differentiable solver produces the corresponding ground-state data $\boldsymbol{x}(\boldsymbol{\theta})$, which the encoder maps to a latent representation $\boldsymbol z (\boldsymbol{\theta}) = E (\boldsymbol{x}(\boldsymbol{\theta}))$. The objective of the pathfinder is to push the latent representation $\boldsymbol{z}(\boldsymbol{\theta})$ away from the centroid $\boldsymbol{\mu}$ of the training cluster.
Concretely, we define the pathfinder loss as,
\begin{equation}
    \mathcal L (\boldsymbol{\theta}) = - \frac{1}{\sqrt{d_z}} \| \boldsymbol z(\boldsymbol{\theta}) - \boldsymbol{\mu} \|_2
    \; ,
\end{equation}
where $d_z$ is the latent dimension. The minus sign is included so that minimizing $\mathcal L$ is equivalent to maximizing the $\ell_2$ distance between the latent representation $\boldsymbol z (\boldsymbol{\theta})$ and the centroid $\boldsymbol{\mu}$. Notice that the loss is a function of the Hamiltonian parameters $\boldsymbol{\theta}$. Because the solver is differentiable, the gradients of the loss can propagate from the latent representation $\boldsymbol z (\boldsymbol{\theta})$ through the encoder and the solver, all the way back to $\boldsymbol{\theta}$. This transports the geometric criterion, distance from the centroid in latent space, into a learning signal on the Hamiltonian parameters, steering $\boldsymbol{\theta}$ toward states farther from the training centroid in latent space.

The full autonomous phase discovery system operates as follows, also shown in Fig.~\ref{fig:schematics}. Given an arbitrary Hamiltonian $\mathcal H (\boldsymbol{\theta})$ and initial starting points $\boldsymbol{\theta}_0$, the Hamiltonian solver is queried to generate a set of training data for AE. The training data are drawn from the vicinity of $\boldsymbol{\theta}_0$ with radius $r_0$. $r_0$ is chosen small enough so that the training data lies within a single phase. The encoder of the trained AE is taken as the embedding model, and the centroid $\boldsymbol{\mu}$ is computed from the latent representations of the training data.
Once the phase detection model is in place, the phase discovery loop begins.
At each step, the solver produces the ground state at the current $\boldsymbol{\theta}$, which the encoder maps to a latent representation $\boldsymbol{z}(\boldsymbol{\theta})$. The pathfinder loss $\mathcal{L}$ is evaluated from $\boldsymbol{z}(\boldsymbol{\theta})$ and $\boldsymbol{\mu}$, and its gradient is backpropagated through the encoder and the solver to $\boldsymbol{\theta}$.
The Hamiltonian parameters are then updated by the gradient descent,
\begin{equation}
    \boldsymbol{\theta} \leftarrow  \boldsymbol{\theta} - \eta \, \nabla_{\boldsymbol{\theta}} \mathcal L
    \; ,
\end{equation}
where $\eta$ is the learning rate.

In the phase discovery loop, the gradient itself carries information about phase structure that can be exploited. As the phase boundary is approached, the spectral gap of the Hamiltonian usually closes, and the eigenvector derivatives scale as $1/\Delta E$ where $\Delta E$ is the spectral gap~\cite{Xie_2020, Wang2019BackpropagationFriendlyE}. The gradients $\nabla_{\boldsymbol{\theta}} \mathcal L$ therefore diverge at gap-closing points, giving an automatic signal that the pathfinder is crossing a phase boundary. This signature is sharpest for diagonalization-based solvers, such as ED and DMRG.
When such a divergence occurs, we cannot continue gradient descent directly. Instead, we revert $\boldsymbol{\theta}$ to the most recent step at which the gradient remains finite, and then take a finite jump in the direction of the last successful update. This ``kick" carries the search across the boundary into the next phase while also detecting the boundary itself.

On the other hand, when the vanishing gradients are encountered, it signals that the search has settled within a region over a sustained window of steps.
The latent representation can no longer be pushed further from the centroid by local moves in $\boldsymbol{\theta}$. We interpret this as the pathfinder having explored the current phase as far as the existing embedding allows. We then retrain the AE on an expanded dataset that now includes a fresh set of ground-state data sampled around the stalled $\boldsymbol{\theta}$. We call this step autoencoder bootstrap. The new model captures both the original phase and the newly discovered region, and the pathfinder resumes from the same point with an updated centroid.

\section{Demonstrations across models and solvers}
\label{sec:num}

\begin{figure*}[!t]
    \centering
    \includegraphics[width=0.95\linewidth]{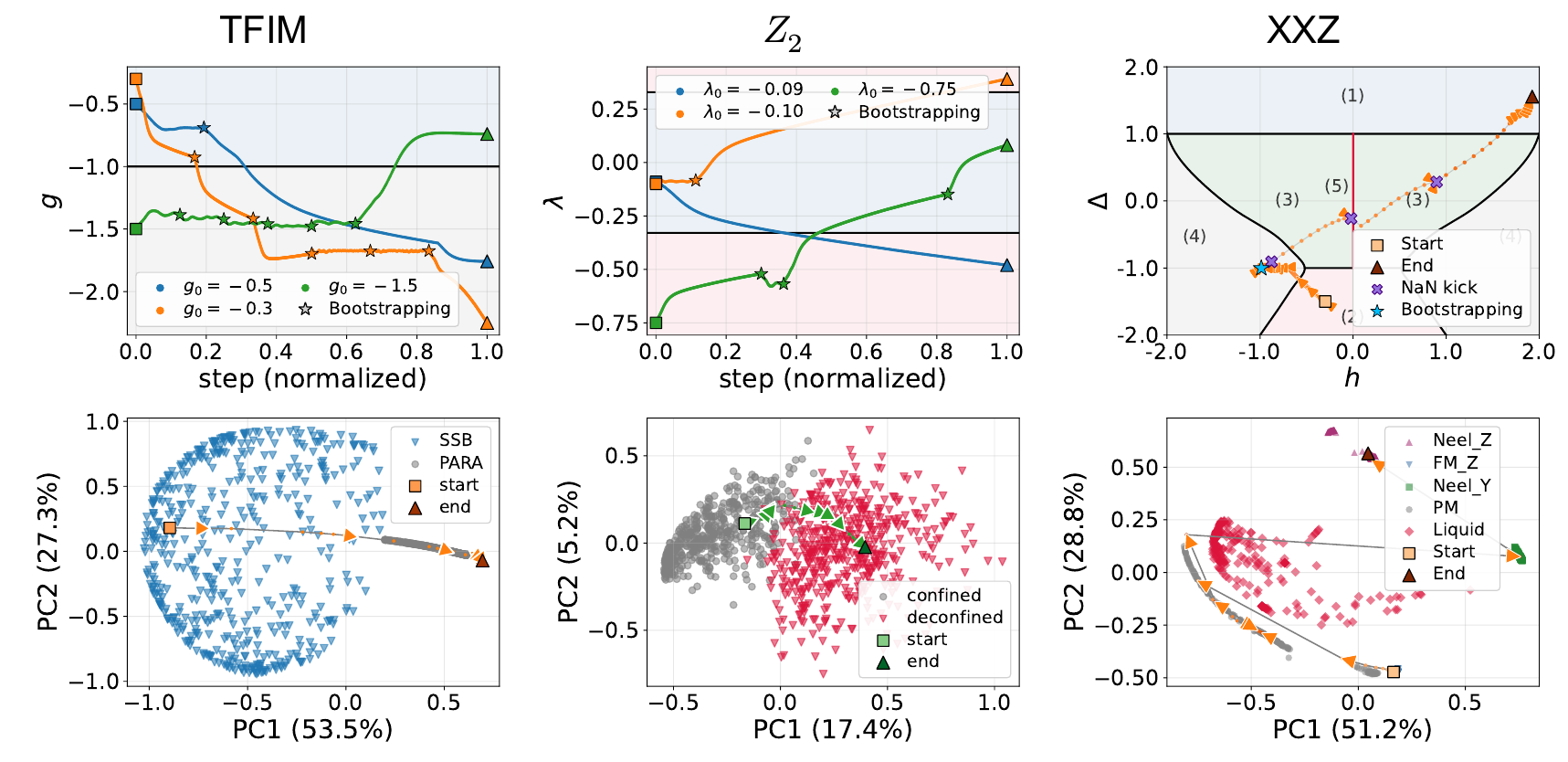}
    \caption{Autonomous phase discovery for TFIM, $\mathbb Z_2$ lattice gauge theory, and the transverse-field XXZ model. Top row: discovery trajectories in the Hamiltonian parameter space; bottom row: the corresponding trajectories projected onto the first two principal components of the autoencoder latent space.
    TFIM: chain of $L=10$ solved by ED. The latent dimension is $d_z = 10$. Three trajectories are shown: $g$ from $-0.3$ to $-2.25$ (orange), from $-0.5$ to $-1.75$ (blue), and from $-1.5$ to $-0.74$ (green), demonstrating crossings of the phase boundary (solid black horizontal line) from either phase to the other. AE bootstrapping is indicated by the star.
    $\mathbb Z_2$: $2 \times 3$ lattice solved by ITE. The latent dimension is $d_z = 30$. Three trajectories are shown: $\lambda$ from $-0.09$ to $-0.48$ (blue, hidden behind the starting point square), from $-0.10$ to $0.39$ (orange), from $-0.75$ to $0.08$ (green). The two trajectories starting in the deconfined phase choose two opposite directions to land in the confined phase.
    XXZ: chain of $L=20$ solved by DMRG. The latent dimension is $d_z = 15$. The starting point is $(h_0, \Delta_0) = (-0.3, -1.5)$ in the $z$-directional ferromagnetic phase labeled by (2), and it traverses the paramagnetic (4), $y$-directional Néel order (3), Luttinger liquid phase (5), finally lands in $z$-directional Néel order at $(h_f, \Delta_f) = (1.92, 1.55)$. Clear and distinct clusters of the five phases are formed in the first two principal components projection of the latent space. The trajectory (orange arrows) moves through each of the clusters.
    Please see Appendix~\ref{app:hyper} for the full hyperparameter list.}
    \label{fig:results}
\end{figure*}

In this section, we demonstrate the system on three models of increasing difficulty, each paired with a different solver: the one-dimensional transverse field Ising model (TFIM), the two-dimensional quantum $\mathbb Z_2$ lattice gauge theory, and the one-dimensional XXZ model with a transverse field. The ground-state information comes from ED on TFIM, ITE on the $\mathbb Z_2$ lattice gauge theory, and DMRG on XXZ. TFIM and the $\mathbb Z_2$ theory have one tuning parameter and two phases each, while the transverse-field XXZ model has two parameters and five distinct phases; our autonomous system traverses all of them.

\subsection{TFIM}
\label{subsec:results_tfim}

We first demonstrate the method on the TFIM,
\begin{equation}
    \mathcal H_{\text{TFIM}} (g) = -\sum_{i=1}^{L-1} \sz_i \sz_{i+1} + g \sum_{i} \sx_i
    \; .
\end{equation}
In the thermodynamic limit, the model is in the spontaneous symmetry breaking (SSB) phase when $|g| < 1$ and the paramagnetic phase when $|g| > 1$. The critical points are exactly at $g = \pm 1$. We use exact diagonalization as the first-principles trial solver for TFIM, albeit its inefficiency. Whenever the solver is queried at a given $g$, whether in the process of training the autoencoder or in the discovery mode, it produces a full wavefunction.

We run our autonomous phase discovery system for TFIM across various starting points in both phases to demonstrate the robustness of our system, $g_0 = -0.3, ~ -0.5$, and $-1.5$, as shown in Fig.~\ref{fig:results} (TFIM). For the runs with $g_0 = -0.3$ and $-0.5$ in the SSB phase, our system identifies the correct direction towards the opposite phase, crosses the phase boundary, and moves deep into the paramagnetic phase. Similar behavior happens in the run with $g_0 = -1.5$ in the paramagnetic phase. Note that apart from the initial training of the autoencoder, the bootstrapping happens at various points in the process of discovery.

The trajectory in the latent space with $g_0 = -0.3$ is shown in the bottom panel of Fig.~\ref{fig:results} (TFIM). For all the TFIM runs, the latent space dimension is chosen to be $d_z = 10$. We perform the principal component analysis of the latent space and select the first two dominant components, which account for $80.8\%$ of the variance in the reference latent representations. We then project the trajectory into the reduced two-component space. The detail of the calculation can be found in Appendix~\ref{app:pca}. The initial point $g_0=-0.3$ sits deep in the SSB cluster and our phase discovery system correctly identifies the direction of the opposite phase and drives the parameter into the paramagnetic regime.

\subsection{$\mathbb Z_2$ lattice gauge model}
\label{subsec:results_z2}

The $\mathbb Z_2$ lattice gauge theory with the following Hamiltonian has two phases: deconfined and confined~\cite{Kitaev_2003, PhysRevD.19.3682, PhysRevD.11.395, Zhou_2021},
\begin{equation}
    \mathcal H_{\mathbb Z_2} (\lambda) = - \sum_{\text p} \prod_{i \in p} \sz_i - \lambda \sum_{i} \sx_i
    \; ,
\end{equation}
In the deconfined phase, corresponding to small $|\lambda|$, the model exhibits a $\mathbb Z_2$ topological order which cannot be characterized by any local order parameters, fundamentally different from the symmetry broken phase in the TFIM. Its ground states are long range entangled and the excitations have anyonic statistics. At large $|\lambda|$, the model is in the trivial confined phase, where the ground state approaches a product state polarized along the field as $|\lambda|\to\infty$, $\ket{+}^{\otimes N}$ for $\lambda>0$ and $\ket{-}^{\otimes N}$ for $\lambda<0$. The critical point between the confined and deconfined phases is at $|\lambda| \approx 0.329$. We choose imaginary time evolution as the solver for this model, which is a deliberate choice to test the versatility of our algorithm with different solvers.

The result is shown in Fig.~\ref{fig:results} ($\mathbb Z_2$). We pick $\lambda_0 = -0.09$, $\lambda_0 = -0.10$ and $\lambda_0 = -0.75$ as our starting points. All three runs are able to cross the phase boundary and find the opposite phase. Notice that $\lambda_0 = -0.09$ and $\lambda_0 = -0.10$ runs take completely opposite trajectories to land in the confined phase. $\lambda_0 = -0.09$ finds the confined phase with negative transverse field, while $\lambda_0 = -0.1$ identifies the direction of confined phase in the positive transverse field region. The search reaches both signs of the field in the confined phase; these regions are related by the unitary transformation $\prod_i\sz_i$, which maps $\lambda$ to $-\lambda$.

Similarly, we show a discovery trajectory with $\lambda_0 = -0.75$ in the latent space. Notice that the first two principal components account for only $22.6\%$ of the variance in the reference latent representations, as opposed to the $80.8\%$ in the TFIM case. This is because the phases in TFIM can be characterized by a simple local order parameter, the magnetization, whereas the topological phase in $\mathbb Z_2$ admits no such local order parameter and is highly non-trivial. Its defining long-range entanglement is spread across many components of the latent space rather than compressed into a few. Furthermore, the transition itself in $\mathbb Z_2$ lattice gauge theory is fundamentally different from that in the TFIM. It is not of the Landau symmetry-breaking type but a continuous confinement-deconfinement transition between a topologically ordered phase and a trivial one. Accordingly, the fact that the two latent-space clusters connect continuously at the phase boundary, rather than being separated by the discrete jump observed for the TFIM, suggests that the autoencoder can capture qualitative differences between distinct types of phase transitions. %
The latent space dimension is chosen to be $d_z = 30$ for $\mathbb Z_2$.

\subsection{XXZ model}
\label{subsec:results_xxz}

Further, we show how well our autonomous phase discovery system performs against a model with two tuning parameters and five distinct phases. Specifically, we use the anisotropic XXZ chain with a transverse field~\cite{Dmitriev_2002, Langari_2004, Langari_2006} for demonstration,
\begin{align}
    \mathcal{H}_{\text{XXZ}}(\Delta, h) = \sum_{i=1}^{L-1} \Big( \sx_i \sx_{i+1} + \sy_i \sy_{i+1} &+ \Delta \sz_i \sz_{i+1} \Big) \nonumber \\
    &+ h \sum_i \sx_i
    \; ,
\end{align}
where $\Delta$ is the anisotropy parameter and $h$ is the transverse field strength.

For TFIM and $\mathbb Z_2$ lattice gauge theory, the parameter space is one-dimensional and contains only two phases, so the system automatically lands the search in a new phase upon exiting the current one, whereas the transverse field XXZ chain is a much more demanding test case. Its parameter space $(\Delta, h)$ is two-dimensional and supports five distinct phases: 1) Néel order along $z$-direction; 2) ferromagnetic order along the $z$ direction; 3) Néel order along $y$-direction; 4) a paramagnetic phase; (5) gapless Luttinger liquid, as shown in Fig.~\ref{fig:results} (XXZ). After crossing a phase boundary, the algorithm needs to choose which direction to continue searching. This ambiguity can be handled by bootstrapping the phase detection model, retraining the AE with an extended set of data consisting of new phase and previous phases. In other words, the retrained model encodes a record of the phases visited by the pathfinder, and the gradient of the pathfinder loss then points toward whichever direction is most ``unfamiliar" to the current embeddings.

We choose DMRG to simulate the transverse field XXZ model, and the ground-state data consist of the local expectation values $\langle \sz_i \rangle$, $\langle \sx_i \rangle$, and $\langle \sy_i \rangle$. These inputs contain only local information and distinguish most of the XXZ phases in this demonstration.
Fig.~\ref{fig:results} (XXZ) shows a trajectory starting with initial parameters $\Delta_0=-1.5$ and $h_0=-0.3$ in the $z$-directional ferromagnetic phase. It crosses into the paramagnetic phase, and triggers an autoencoder bootstrapping (cyan star). With an updated latent space containing both the ferromagnetic and paramagnetic phase, the pathfinder is able to identify a new direction and carries on the search for new phases.

Remarkably, the gapless Luttinger liquid phase is also identified by our system, indicated by the purple marker near the $h=0$ line (red) in Fig.~\ref{fig:results} (XXZ). As the pathfinder approaches this region, the gap between the ground and first excited states closes, which causes the gradients of loss to diverge and triggers the ``kick". Because the Luttinger liquid occupies the $h=0$ line rather than a finite area of the $(\Delta, h)$ plane, the pathfinder cannot settle inside it. The phase therefore registers along the trajectory as a gradient-divergence event rather than as a region in which the search comes to rest. The autonomous phase discovery system has no built-in notion of gaplessness, yet gap closing is a textbook signature of quantum criticality and of many gapless phases of interest. What our algorithm recovers, in a fully unsupervised manner, is precisely this hallmark: the divergence of the gradient is the closing of the gap, expressed at the level of the optimization dynamics.

After the $y$-directional Néel phase and the gapless Luttinger liquid, the algorithm finally lands in the $z$-directional Néel phase. By the end, it is able to find all five phases in the transverse-field XXZ model. The trajectory in the latent space is shown in the bottom panel of Fig.~\ref{fig:results} (XXZ). The latent space dimension for XXZ is chosen to be $d_z = 15$. The first two principal components account for $80\%$ of the variance in the reference latent representations. The discovery path starts in the cluster of the $z$-directional ferromagnetic phase (blue triangle), moves toward the paramagnetic line (gray circle), passes by the liquid phase, and lands at the $y$-directional (green square) and $z$-directional (purple triangle) Néel orders, consistent with the discovery trajectory in the parameter space. Notice that the Luttinger liquid states in the reference set form the most diffuse cluster in the projection, while the gapped phases form much tighter clusters.

\section{Phase discovery at scale}
\label{sec:scale}

The demonstrations presented in Sec.~\ref{sec:num} operate in parameter spaces of at most two dimensions, with at most five distinct phases. In this section, we further test our autonomous system on free-fermionic models with significantly larger numbers of parameters and phases: the generalized cluster-state chain with up to $d=200$ phases where $d$ denotes both the number of coupling parameters and the number of phases, and a $50$-parameter Chern insulator whose number of accessible phases is not known in advance. Specifically, we compare the performance of our system with random sampling on these two free-fermionic models, using the number of phases covered as the metric. We show that our system outperforms the random sampling, with the coverage advantage increasing with coupling dimension in the cluster benchmark and with query budget in the Chern benchmark.

\subsection{Protocol}
\label{subsec:results_protocol}

Given a Hamiltonian parameterized by $\boldsymbol{\theta} \in \mathbb R^{|\boldsymbol{\theta}|}$, where $|\boldsymbol{\theta}|$ denotes the dimension of the parameter space, random sampling draws parameter vectors from $\mathbb R^{|\boldsymbol{\theta}|}$ according to the rotation-invariant Gaussian measure. We determine the number of phases covered by the random sampling using the winding number or Chern number, respectively, and compare it with the number covered by our autonomous phase-discovery system at an identical query budget, where each query corresponds to one ground-state evaluation. %

Random sampling is a meaningful baseline to compare to when the number of phases is large and the phases occupy regions of widely varying sizes in a high-dimensional parameter space. If a phase sector has probability mass $v$ under the sampling distribution, the probability that it is missed by $N$ independent random samples is $(1-v)^N$. Random sampling therefore becomes unlikely to resolve sectors with $v\ll 1/N$.
In contrast, a trajectory-based search can enter a small phase sector by crossing into it from a neighboring region, such that its probability mass under the sampling distribution is not necessarily the limiting factor. Our benchmarks show that the autonomous discovery system can indeed identify small phase sectors that are missed by random sampling at the same query budget.

Random sampling also requires no prior knowledge beyond the parameterization, whereas Bayesian optimization and active learning~\cite{Kusne_2020,settles2009active} require an acquisition objective or feedback that is not naturally available in the label-free phase-discovery setting. One important distinction is that random sampling alone provides no mechanism for identifying the phases represented by the sampled states, whereas our autonomous system integrates phase detection directly into the search and uses the detected phase structure to guide subsequent exploration. In principle, random sampling could be supplemented with an unsupervised phase-detection method as a post-processing step. Since our comparison focuses on phase coverage rather than phase identification, we do not include such a post-processing step. Instead, the winding and Chern numbers are used only for evaluation to determine the number of distinct phases covered by each method.

\subsection{Cluster chain with up to 200 phases}
\label{subsec:results_bdi}

\begin{figure}[t]
    \centering
    \includegraphics[width=0.95\columnwidth]{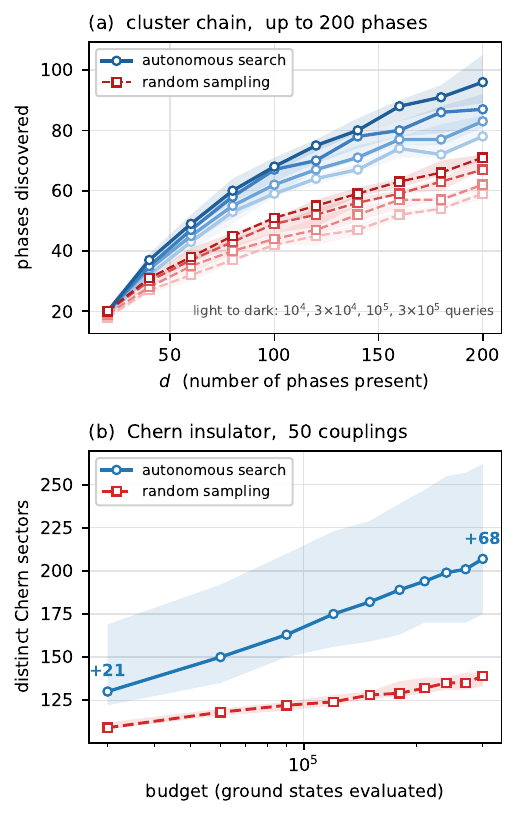}
    \caption{Phases discovered by the autonomous search versus random sampling at equal numbers of ground-state evaluations. (a)~Generalized cluster chain, Eq.~\eqref{eq:cluster}, at $L=500$: phases found across $d\in[20,200]$, at budgets of $10^4$, $3\times10^4$, $10^5$, and $3\times10^5$ evaluations (light to dark). In 36 of the 40 $(d,\text{budget})$ cells the search's worst seed exceeds the baseline's best realization. (b)~Chern insulator, Eq.~\eqref{eq:chernmodel}: distinct Chern sectors against budget; every budget shown is a nested prefix of the same runs, and the median margin grows from $+21$ at $3\times10^4$ to $+68$ at $3\times10^5$, increasing end to end in every seed. Medians over five fresh seeds throughout; bands are min--max over seeds or realizations.}
    \label{fig:scale}
\end{figure}

The first scaling benchmark uses the generalized cluster-state chain,
\begin{align}
    \mathcal H_{\text{cluster}} (\boldsymbol{t}) =
        &- t_0 \sum_{i=1}^{L} \sx_i
        - t_1 \sum_{i=1}^{L-1} \sz_i \sz_{i+1}
        \nonumber \\
        &- \sum_{\alpha=2}^{d-1} t_{\alpha}
        \sum_{i=1}^{L-\alpha}
        \sz_i \sx_{i+1} \dots \sx_{i+\alpha-1} \sz_{i+\alpha}
        \; ,
    \label{eq:cluster}
\end{align}
where $d$ sets both the number of couplings $\boldsymbol t=(t_0,\dots,t_{d-1})$ and the number of distinct phases, characterized by winding numbers $\omega=0,\dots,d-1$. Under open boundary conditions, a phase with winding number $\omega=\alpha$ hosts $\alpha$ unpaired Majorana modes at each edge. The benchmark itself uses a closed chain, with site indices taken modulo $L$, and is restricted to the even-parity sector, corresponding to antiperiodic boundary conditions for the fermions. This family therefore provides a controlled setting in which both the parameter-space dimension and the number of phases can be systematically increased, up to $d=200$ in our experiments.

We work in the free-fermion representation. Under the Jordan--Wigner transformation, Eq.~\eqref{eq:cluster} maps to a long-range Kitaev chain~\cite{kitaev2001} in symmetry class BDI~\cite{verresen2017},
\begin{equation}
    \mathcal H =
    -\sum_{\alpha=0}^{d-1}c_\alpha\sum_j i\,b_j a_{j+\alpha},
    \qquad
    c_0=-t_0,\quad c_{\alpha\geq1}=t_\alpha .
    \label{eq:majorana}
\end{equation}
For the closed chain used in the benchmark, the phase is determined by the winding number of
$f(k)=\sum_\alpha c_\alpha e^{ik\alpha}$, which can be obtained exactly by counting the roots of $\sum_\alpha c_\alpha z^\alpha$ inside the unit circle. The ground state is fully specified by the Majorana correlations
\begin{equation}
    g(r)=\frac{1}{L}\sum_k
    \frac{f(k)}{|f(k)|}e^{-ikr},
    \label{eq:polar}
\end{equation}
which form the input to the encoder. Importantly, the winding number is not provided to the autonomous search and is used only afterward to evaluate the number of distinct phases covered. The physical derivation is given in Appendix~\ref{app:majorana-covariance}.

Fig.~\ref{fig:scale} (a) compares the phase coverage of the autonomous search and random sampling at identical ground-state-evaluation budgets, for $d$ ranging from $20$ to $200$ for a chain of length $L=500$. For each budget, we run the search five times from a fixed starting point, varying only the autoencoder's random seed, and report the median over the five runs. At $d=20$, both methods saturate all available phases. As $d$ increases, however, the autonomous search systematically covers more phases than random sampling, with a gap of $5$--$25$ phases for $d\geq40$. The advantage increases with $d$ and persists across budgets from $10^4$ to $3\times10^5$ evaluations. Remarkably, the autonomous search at $10^4$ evaluations already exceeds the phase coverage reached by random sampling at $3\times10^5$, demonstrating that the advantage reflects improved search efficiency rather than simply access to a larger evaluation budget.

The growing advantage can be traced in part to phase sectors with very small probability under the random-sampling distribution. At $d=140$, for example, the union of four autonomous-search seeds at a budget of $10^5$ contains $21$ winding sectors that are not encountered in four million random samples. The autonomous search can reach such sectors by traversing neighboring regions of parameter space rather than relying on directly sampling them.

\subsection{A 50-parameter Chern insulator}
\label{subsec:results_chern}

The second scaling benchmark is designed to test a more challenging setting in which the number and distribution of accessible phases are not known in advance. Specifically, we generalize the Qi--Wu--Zhang (QWZ) model~\cite{qwz2006} by extending the hoppings beyond nearest neighbor,
\begin{align}
    d_x &= \sum_{n=1}^{H} a_n \sin(n k_x)
    \;\; ; \;\;
    d_y = \sum_{n=1}^{H} b_n \sin(n k_y)
    \; , \nonumber\\
    d_z &= m + \sum_{n=1}^{H} c_n \big( \cos(n k_x) + \cos(n k_y) \big) \nonumber\\
    &\quad + e \, \cos(k_x) \cos(k_y)
    \; .
    \label{eq:chernmodel}
\end{align}
This is a two-band Hamiltonian $\mathcal H(\mathbf{k})=\mathbf{d}\cdot\boldsymbol\sigma$ with $H=16$, giving a $50$-parameter family, where parameters are $\boldsymbol{\theta} = (a_n, b_n, c_n, e, m)$. The additional hopping terms provide a simple way to increase the dimensionality of the parameter space and allow a broad range of topological sectors, labeled by the Chern number $\mathcal C$~\cite{tknn1982,haldane1988}.

Unlike the generalized cluster chain, where both the number of phases and explicit representatives of each phase are known in advance, the generalized QWZ model has no known enumeration of its accessible Chern sectors or general construction for realizing a prescribed Chern number. Thus, while evaluating $\mathcal C(\boldsymbol{\theta})$ for a given parameter point $\boldsymbol{\theta}$ is straightforward, the inverse problem of constructing a parameter point with a prescribed $\mathcal C$ is not. We discuss this inverse problem, bounds on the accessible Chern numbers, and simple harmonic constructions in Appendix~\ref{app:inverse}.

For each parameter point, the autoencoder receives the normalized Bloch vector $\hat{\mathbf d}(\mathbf k)=\mathbf d(\mathbf k)/|\mathbf d(\mathbf k)|$ sampled on a $32\times32$ momentum-space grid, which provides a compact representation of the occupied band and can be computed directly from Eq.~\eqref{eq:chernmodel} without diagonalization. The Chern number is not provided to the autoencoder and is used only to evaluate phase coverage. We compute it separately using the gauge-invariant lattice method of Fukui, Hatsugai, and Suzuki~\cite{fhs2005} on a finer $768\times768$ grid. See Appendix~\ref{app:bench} for details.

Fig.~\ref{fig:scale} (b) shows the equal-budget comparison between the autonomous search and random sampling, with the autonomous search initialized at a gapped mass-only point with $\mathcal C=0$. At $3\times10^5$ ground-state evaluations, the autonomous search discovers a median of $207$ distinct Chern sectors across five seeds, compared with $139$ for random sampling. The performance gap also increases with the query budget in every seed. Taking the union of the sectors found across all five runs, the autonomous search reaches $283$ distinct sectors spanning $\mathcal C\in[-192,188]$, while random sampling reaches $160$ sectors at the same total budget of $1.5\times10^6$ evaluations. All $160$ sectors found by random sampling are also found by the autonomous search. Even with an additional $10^6$ samples, random sampling reaches only $168$ sectors, compared with the $283$ found by the autonomous search.

\begin{figure}[t]
    \centering
    \includegraphics[width=0.95\columnwidth]{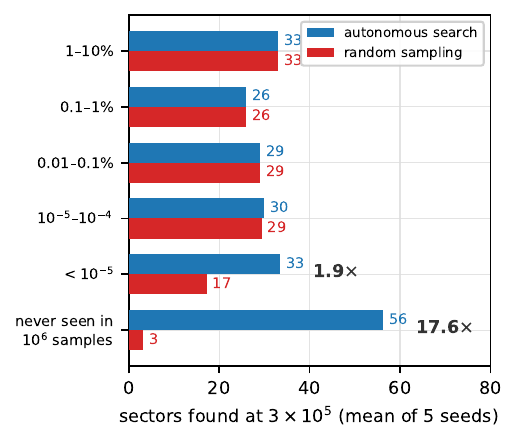}
    \caption{The comparison of Fig.~\ref{fig:scale}(b) at $3\times10^5$ evaluations, resolved by the probability mass occupied by each Chern sector under the Gaussian sampling measure (mean over five seeds). The four decades resolved by the $10^6$-sample reference are saturated by both methods, while the advantage of the autonomous search is concentrated in the two rarest bins.}
    \label{fig:chern_vol}
\end{figure}

Fig.~\ref{fig:chern_vol} further compares the two methods based on the probability mass of each Chern sector under the random-sampling distribution. The two methods find similar numbers of sectors in regions that are well sampled by random sampling. The advantage of the autonomous search comes mainly from rare sectors with probability mass below $10^{-5}$. This suggests that the trajectory-based search is particularly effective at reaching small phase regions that are unlikely to be found by independent random samples.

\section{Discussion \& Outlook}
\label{sec:outlook}

We have introduced a fully autonomous system for quantum phase discovery that recasts the search for new phases in the Hamiltonian parameter space as a geometric problem in the latent space of an autoencoder, navigated by gradient descent through a differentiable solver, without supplying phase labels or order parameters to the search, and without a predetermined parameter sweep.

Different phases produce qualitatively different signatures, and a procedure sensitive to only one of them will miss whole classes. Symmetry-broken phases form clusters in the latent space; the gapless Luttinger liquid has no local order parameter and zero measure in the phase diagram, and is detected instead through the divergence of the gradient. The system reads both signals, which is why it works across the models tested here.

Whether sampling or a walk is preferable depends on the geometry of the target sector, in particular its volume and its connectivity to the rest of the diagram. Uniform sampling must land inside a sector, so a sector with probability mass $v$ under the sampling distribution requires of order $1/v$ independent draws, and uninformed random sampling is therefore intrinsically inefficient for the rare sectors. A trajectory can enter a rare sector from a neighboring region, although detecting it still requires evaluating a point inside it. In exchange, the walk is restricted to phases connected to the one it starts in. Because the two fail for different reasons, combining them is worth trying: a random pass to distribute starting points across the diagram, and a walk from each working outward into the rare regions. We have not tested this.

Reaching sectors missed by random sampling at the tested budgets can still require many ground-state evaluations. Averaged over a run it spends on the order of a thousand ground-state evaluations for every sector it adds, which is to say that almost every state it computes belongs to a sector already visited. One possible limitation is that the repulsion acts on visited points rather than on entire phase regions. Training the recognizer once on a broad random pass and then holding it fixed would measure novelty against every phase seen so far rather than against the cloud around the current point, and would remove the retraining that now interrupts the walk; the random baseline computed for the comparison is already such a pass. Systems with many complex phases may require further upgrades to the loss, such as maximizing the reconstruction error or representing each phase by several centroids. Where a label can be obtained the repulsion can act on sectors instead, and the search can be aimed at a prescribed invariant. Such a label-aware search should also be compared with methods that use the same label information, including active learning~\cite{settles2009active,Kusne_2020}.

Each of these limitations suggests a direction for future work. Replacing the differentiable solvers with neural-network-based solvers that are differentiable by construction~\cite{fitzek2024rydberggpt, lange2024architecturesapplicationsreviewneural} would relax the cost and size constraints of exact diagonalization, density-matrix renormalization group, and imaginary-time evolution, and could reduce the cost of gradient-based navigation of parameter space. A natural further test is to apply the system to interacting fermionic models, where two-dimensional gapless phases would probe the recognition layer in the regime that is hardest for conventional methods, and where the label must be kept discrete and exactly computable: without one the count on which the whole comparison rests is no longer defined. Finally, the system should be stress-tested against tuning parameters that are not restricted to coupling strengths in the Hamiltonian, such as lattice geometry or the terms present in the Hamiltonian at all.

\begin{acknowledgments}
We thank Juan Carrasquilla for helpful discussion on machine learning for phase classification; the last author thanks Lei Wang and Jinguo Liu for helpful discussion on differentiable solvers. S.Z. thanks Y.Z. and X.L. for their support through pregnancy and new parenthood over the course of this project, which took roughly twice as long as Zoryana Z., now one year old, has been alive. S.Z. thanks Subir Sachdev for hosting her at Harvard Univerisy. Claude Code and Codex was used to assist with portions of the code development. Research at Perimeter Institute is supported in part by the Government of Canada through the Department of Innovation, Science and Industry Canada and by the Province of Ontario through the Ministry of Colleges and Universities. Y.Z. is supported by funding from the European Research Council (ERC) under the European Union's Horizon 2020 research and innovation programme (grant agreement No.~864597)
\end{acknowledgments}

\bibliography{references}

\appendix
\onecolumngrid

\section{Principal Component Analysis of the Latent Space}
\label{app:pca}

The latent spaces used in this work have dimension $d_z \gg 2$, so the latent-space trajectories in Fig.~\ref{fig:results} are visualized by projecting onto the two dominant principal components of a reference set of latent representations. Let $\{ \boldsymbol z_i \}_{i=1}^{n}$ denote the latent representations of the reference data produced by the trained encoder, and let $\bar{\boldsymbol z} = \frac{1}{n} \sum_i \boldsymbol z_i$ be their mean. We assemble the centered data matrix $\tilde Z \in \mathbb R^{n \times d_z}$ with rows $\boldsymbol z_i - \bar{\boldsymbol z}$ and compute its singular value decomposition,
\begin{equation}
    \tilde Z = U \, \Sigma \, V^{\mathsf T}
    \; ,
\end{equation}
where the columns $\boldsymbol v_k$ of $V$ are the principal axes and the singular values $\sigma_1 \geq \sigma_2 \geq \dots$ measure the variance captured by each axis. Any latent representation $\boldsymbol z$, including those of the discovery trajectory, is then projected as
\begin{equation}
    \boldsymbol y = \big( \boldsymbol v_1 \cdot (\boldsymbol z - \bar{\boldsymbol z}), \; \boldsymbol v_2 \cdot (\boldsymbol z - \bar{\boldsymbol z}) \big)
    \; ,
\end{equation}
and the fraction of the embedding information retained by component $k$ is quantified by the explained variance ratio $\mathrm{EVR}_k = \sigma_k^2 / \sum_j \sigma_j^2$. The percentages quoted in the main text are $\mathrm{EVR}_1 + \mathrm{EVR}_2$ of the corresponding reference sets.

The reference sets are the training data of the final machine learning model of each experiment: ground states sampled from both phases for the TFIM; $500$ confined and $500$ deconfined ground states for the $\mathbb Z_2$ lattice gauge theory; and the all-phases training set for the XXZ model. The projection axes are fixed by the reference set alone, and the discovery trajectories are only projected onto them.

\section{Hyperparameters}
\label{app:hyper}

Every run in this work shares the same architecture family and discovery loop. The machine learning is a fully connected autoencoder with ReLU activations whose encoder output is normalized onto the unit hypersphere; it is trained full-batch with AdamW. Wavefunction inputs (TFIM, $\mathbb Z_2$) use the infidelity loss $\mathcal L_{\text C} = 1 - |\bra{\phi}\ket{\psi}|^2$, while the observable-vector input (XXZ) uses the mean squared error. In the discovery loop, a diverging gradient triggers the jump: the learning rate $\eta$ is temporarily boosted and the parameters are kicked along the last successful update direction; a stalled search (vanishing gradient or vanishing parameter change) triggers the autoencoder bootstrap, retraining the AE on the union of the previous training data and a fresh circle of samples around the stalled point. Table~\ref{tab:hyper} lists the hyperparameters and run configuration of each.

A few per-run details do not tabulate well. The $\mathbb Z_2$ runs pin the imaginary time evolution to a fixed branch of the four-fold degenerate ground manifold. The XXZ embedding centroid is history aware: it averages the circle centroids of all visited phases. The generalized-cluster-state couplings $\boldsymbol t$ are normalized onto the unit sphere (samples are projected back after the Gaussian draw), the chain has periodic boundary conditions in the even-parity sector, the pathfinder rate is cosine-decayed within each block with the gradient clipped at global norm $0.5$, and the run terminates once a whole block moves $\boldsymbol t$ by less than $10^{-4}$. In the demonstrations of Sec.~\ref{sec:num} the jump also kicks the parameters by twice the last successful update; the long-horizon benchmark loop has no jump.

\begin{table*}[!t]
    \centering\footnotesize
    \setlength{\tabcolsep}{3.4pt}
    \begin{tabular}{@{}l l l l l l@{}}
        \toprule
        & TFIM & $\mathbb Z_2$ & XXZ & cluster chain & Chern \\
        \midrule
        Solver & ED & ITE ($d\tau{=}10^{-2}$, 150 st.) & DMRG ($\chi{=}10$, 10 sw.) & closed form & closed form \\
        System & $L = 10$ & $2 \times 3$ & $L = 20$ & $L{=}500$, $d{\leq}200$ & $H{=}16$, 50 couplings \\
        Starting point & $g_0 = -0.3, -0.5, -1.5$ & $\lambda_0 = -0.09, -0.1, -0.75$ & $(\Delta_0, h_0){=}(-1.5, -0.3)$ & $e_0$ or random & mass only \\
        AE input (dim.) & $\ket{\psi}$ ($1024$) & $\ket{\psi}$ ($4096$) & $\langle \sigma^{x,y,z}_i \rangle$ ($60$) & $g(r)$ ($500$) & $\hat{\mathbf d}(\mathbf k)$, $32^2$ ($3072$) \\
        Encoder widths & $1024$--$500$--$d_z$ & $4096$--$500$--$d_z$ & $60$--$20$--$d_z$ & $500$--$32$--$d_z$ & $3072$--$32$--$d_z$ \\
        Latent dimension $d_z$ & $10$ & $30$ & $15$ & $16$ & $16$ \\
        AE loss & fidelity & fidelity & MSE & fidelity & fidelity \\
        AE epochs & $10^4$ & $5 \times 10^4$ & $5 \times 10^4$ & $800$ & $1.5 \times 10^3$ \\
        AE learning rate & $10^{-4}$ & $10^{-4}$ & $10^{-5}$ & $10^{-3}$ & $10^{-3}$ \\
        Samples (init/boot.) & $1000$ / $500$ & $500$ / $250$ & $500$ / $500$ & $30$ / $30$ & $30$ / $30$ \\
        Sampling radius $r_0$ & $0.3$ / $0.1$ & $0.05$ & $0.01$ & $0.1$ & $0.087$ \\
        Pathfinder rate $\eta$ & $10^{-2}$ & $5 \times 10^{-3}$ & $10^{-3}$ & $5\times10^{-2}$ & $1.24$, cos. decay \\
        Jump boost ($\times\eta$, steps) & $10$, $5$ & $10$, $5$ & $10^2$, $10$ & --- & --- \\
        Bootstrap trigger & $|\nabla \mathcal L| < 10^{-4}$ & $|\nabla \mathcal L| < 5 \times 10^{-3}$ & $|\Delta \boldsymbol{\theta}| < 10^{-4}$ & block stall & block stall \\
        Stall window (steps) & $25$ & $25$ & -- & $20$ & $20$ \\
        Steps per block & $200 \times 5$ & $1000 \times 5$ & $2000 \times 6$ & $500$ & $319$ \\
        \midrule
        Trail repulsion $w$/$s$/len. & --- & --- & --- & $0.5$ / $0.15$ / $400$ & $0.73$ / $0.083$ / $114$ \\
        Gradient refresh $k$ & --- & --- & --- & $30$ & $5$ \\
        Gradient clip & --- & --- & --- & $0.5$ & $1.008$ \\
        \bottomrule
    \end{tabular}
    \caption{Hyperparameters and run configuration of every run in this work: the demonstrations of Sec.~\ref{sec:num} and the equal-budget benchmarks of Sec.~\ref{sec:scale}. The jump boost row gives the temporary learning-rate multiplier and the number of steps it is applied for; the bootstrap trigger is an EMA-smoothed gradient norm for TFIM and $\mathbb Z_2$ and the maximum parameter change for XXZ, sustained over the stall window. The last three rows apply only to the long-horizon loop of Appendix~\ref{app:bench}. The two benchmark families share that loop but were tuned separately, and the Chern column is the configuration returned by the successive-halving bracket.}
    \label{tab:hyper}
\end{table*}

\section{Benchmark configuration}
\label{app:bench}

The searches of Sec.~\ref{sec:scale} run for as many as $3\times10^5$ queries, two to three orders of magnitude longer than the demonstrations of Sec.~\ref{sec:num}, and three modifications to the loop of Sec.~\ref{sec:algo} keep them productive over that horizon. The pathfinder loss of Sec.~\ref{sec:algo} is modified in three ways: the distance to the centroid is squared and the $1/\sqrt{d_z}$ normalization dropped, and a repulsion from the trajectory's recent history is added,
\begin{align}
    \mathcal L(\boldsymbol\theta) ={}& -\big\|\boldsymbol z(\boldsymbol\theta)-\boldsymbol\mu\big\|_2^2 \nonumber\\
    &+ w \sum_{i\in\text{trail}} \exp\!\left(-\frac{\|\boldsymbol\theta-\boldsymbol\theta_i\|_2^2}{s^2}\right)
    \; ,
    \label{aeq:benchloss}
\end{align}
where the trail holds the most recent solved points, so the walk avoids where it has been rather than only the starting cluster. The parameters are updated with Adam under a cosine-decayed step size, clipped at a global norm, and projected back onto the unit sphere after every step. The gradient is refreshed once every $k$ parameter steps and reused in between; only the refresh points are solved, labeled, and charged against the budget. A stalled block triggers the bootstrap of Sec.~\ref{sec:algo}, retraining the autoencoder around the current point.

Hyperparameters were selected by a successive-halving bracket~\cite{hyperband2018} ($27\to9\to3\to1$ trials, elimination factor $3$) over eight parameters drawn log-uniformly; because a search is continuable, promoted trials resume rather than restart. The bracket ranks trials by a single noisy run each, which biases the winner upward, so the three finalists were re-evaluated on held-out seeds and one of them dropped from $+9$ in-bracket to $-1$; every number in Sec.~\ref{sec:scale} comes from seed families never used in selection. The Chern column of Table~\ref{tab:hyper} lists the winning configuration. The autoencoder input for the Chern family is the normalized field $\hat{\mathbf d}(\mathbf k)$ on a $32\times32$ Brillouin-zone grid; at $H=16$ this grid sits at the Nyquist limit of the highest harmonic, so two of the fifty couplings are invisible to the objective, a defect discussed in Sec.~\ref{sec:outlook}.

The lattice Chern number of Sec.~\ref{subsec:results_chern} is evaluated with the gauge-invariant plaquette construction of Ref.~\cite{fhs2005}, which returns an integer for any grid; what resolving the Berry flux through every plaquette buys is convergence of that integer to the Chern number. At $H=16$ the field $\mathbf d(\mathbf k)$ supports up to $4H^2=1024$ monopole crossings, and resolving them takes far more than the Nyquist rate of the harmonics: on a $96^2$ grid the integer agrees with the $1536^2$ reference on only about a third of random points, with a mean error of $3$ units biased toward small $|\mathcal C|$; agreement reaches $70\%$ at $192^2$, $93\%$ at $384^2$, and $98.2\%$ at $768^2$. All labels in Sec.~\ref{subsec:results_chern} use the $768^2$ grid. The agreement rate at $768^2$ is the same, within a fraction of a percent, on the points visited by the search and on the baseline's Gaussian draws, so the residual error is symmetric between the two methods.

\section{Majorana ground-state correlations and phase characterization}
\label{app:majorana-covariance}

The generalized cluster-chain benchmark admits an efficient free-fermion description that serves two distinct purposes. First, its phases can be labeled exactly by an integer bulk winding number, providing a ground-truth label used only to evaluate phase coverage. Second, its Gaussian ground state is completely specified by a real vector of Majorana two-point correlations, which is used as the input to the autoencoder. Here we derive both results and explain their relation to the Majorana edge-mode interpretation of the phases.

\subsection{From the cluster chain to a long-range Majorana chain}

The Jordan--Wigner Majorana operators
\begin{equation}
a_j=\left(\prod_{\ell<j}\sx_\ell\right)\sz_j,
\qquad
b_j=\left(\prod_{\ell<j}\sx_\ell\right)\sy_j
\end{equation}
are Hermitian and satisfy
\begin{equation}
\{a_j,a_\ell\}=\{b_j,b_\ell\}=2\delta_{j\ell},
\qquad
\{a_j,b_\ell\}=0.
\end{equation}
In particular, $a_j^2=b_j^2=1$. Their bilinears reproduce the spin operators,
\begin{equation}
i b_j a_j=-\sx_j,
\qquad
i b_j a_{j+\alpha}
=
\sz_j
\left(\prod_{s=1}^{\alpha-1}\sx_{j+s}\right)
\sz_{j+\alpha},
\qquad \alpha\geq1.
\end{equation}
Thus the field term has the opposite conversion sign from the longer strings, and Eq.~\eqref{eq:cluster} becomes
\begin{equation}
\mathcal H
=
-i\sum_{j,\ell}b_j C_{j\ell}a_\ell
=
-\sum_{j,\alpha}c_\alpha i b_j a_{j+\alpha},
\qquad
c_0=-t_0,\quad c_{\alpha\geq1}=t_\alpha .
\label{aeq:majorana-C}
\end{equation}
This is a long-range Kitaev chain in symmetry class BDI~\cite{kitaev2001,verresen2017}.

For the closed chain used in the benchmark, we work in the antiperiodic fermion sector,
\begin{equation}
a_{j+L}=-a_j,\qquad b_{j+L}=-b_j,
\end{equation}
so that $C_{j,j+\alpha}=c_\alpha$, with an additional minus sign whenever the index crosses the boundary. The resulting matrix $C$ is a twisted circulant.

\subsection{Bulk winding number and Majorana edge modes}

For translation-invariant couplings, introduce
\begin{equation}
f(k)=\sum_{\alpha=0}^{d-1}c_\alpha e^{ik\alpha}.
\label{aeq:f-k}
\end{equation}
The gapped phases of the closed chain are characterized by the integer winding number
\begin{equation}
\omega
=
\frac{1}{2\pi i}
\int_0^{2\pi} dk\,
\partial_k \log f(k),
\label{aeq:winding}
\end{equation}
provided $f(k)\neq0$ throughout the Brillouin zone. Geometrically, $\omega$ counts how many times the curve traced by $f(k)$ winds around the origin as $k$ traverses the Brillouin zone.

The connection to the open-chain interpretation is particularly transparent along the coupling axes. If only $c_\alpha$ is nonzero,
\begin{equation}
f(k)=c_\alpha e^{i\alpha k},
\end{equation}
and hence $\omega=\alpha$. In real space, the Hamiltonian then couples $b_j$ to $a_{j+\alpha}$. On an open chain this shifted pairing leaves $\alpha$ unpaired Majorana modes at each edge. Thus the bulk winding number of the closed chain and the number of boundary Majorana modes of the corresponding open chain are related by the bulk--boundary correspondence,
\begin{equation}
\omega=\alpha
\quad\Longleftrightarrow\quad
\alpha\ {\rm unpaired\ Majorana\ modes\ per\ edge},
\end{equation}
for the convention used here. The $d$ coupling axes $c=e_\alpha$, $\alpha=0,\ldots,d-1$, therefore provide simple representatives of the $d$ phases with winding numbers $\omega=0,\ldots,d-1$.

For the benchmark, the winding number can be evaluated without discretizing the Brillouin zone. Writing
\begin{equation}
p(z)=\sum_{\alpha=0}^{d-1}c_\alpha z^\alpha,
\end{equation}
we have $f(k)=p(e^{ik})$. By the argument principle, the winding number equals the number of roots of $p(z)$ inside the unit circle, as long as no root lies on the unit circle. We obtain these roots from the companion matrix and therefore determine the phase label without a momentum-space grid or a corresponding discretization error. These winding labels are used only to evaluate the number of phases covered by the different search strategies; they are not supplied to the autoencoder or to the autonomous search.

\subsection{Ground-state correlations from the polar factor}

We next show how the ground state follows directly from the coupling matrix $C$. Take a real singular-value decomposition
\begin{equation}
C=U\Sigma V^{\mathsf T}
\end{equation}
and define
\begin{equation}
\widetilde{\boldsymbol b}=U^{\mathsf T}\boldsymbol b,
\qquad
\widetilde{\boldsymbol a}=V^{\mathsf T}\boldsymbol a.
\end{equation}
Orthogonality preserves the Majorana anticommutation relations. Introducing
\begin{equation}
\eta_\nu=
\frac{\widetilde b_\nu+i\widetilde a_\nu}{2},
\end{equation}
the Hamiltonian becomes
\begin{equation}
\mathcal H
=
-\sum_{\nu=1}^{L}\sigma_\nu
i\widetilde b_\nu\widetilde a_\nu
=
\sum_{\nu=1}^{L}\sigma_\nu
(1-2\eta_\nu^\dagger\eta_\nu),
\qquad \sigma_\nu>0.
\label{aeq:majorana-modes}
\end{equation}
The ground state occupies each $\eta_\nu$ mode in this convention, giving
\begin{equation}
\langle i\widetilde b_\nu\widetilde a_\mu\rangle
=
\delta_{\nu\mu}.
\end{equation}
The singular values $\sigma_\nu$ determine the excitation energies $2\sigma_\nu$, whereas the singular vectors determine the ground-state correlations. Transforming back,
\begin{equation}
G_{j\ell}
\equiv
\langle i b_j a_\ell\rangle
=
(UV^{\mathsf T})_{j\ell},
\qquad
G=C(C^{\mathsf T}C)^{-1/2}.
\label{aeq:majorana-polar}
\end{equation}
Thus the ground-state correlation matrix is the orthogonal polar factor of $C$: the positive singular values, which set the energy scales, are flattened to unity while their eigenvectors are retained.

This expression assumes that $C$ is nonsingular. At a gap closing, where a singular value vanishes, the occupation of the corresponding zero mode must be specified separately and the ground state is not uniquely selected by Eq.~\eqref{aeq:majorana-polar}.

\subsection{Momentum-space form and the correlation vector $g(r)$}

Because $C$ is a twisted circulant, its normalized eigenvectors are
\begin{equation}
v_k(j)=L^{-1/2}e^{ikj}
\end{equation}
at the antiperiodic momenta
\begin{equation}
k_n=\frac{(2n+1)\pi}{L},
\qquad n=0,\ldots,L-1,
\end{equation}
with
\begin{equation}
Cv_k=f(k)v_k.
\end{equation}
Since the coefficients $c_\alpha$ are real,
$f(-k)=f(k)^*$. The eigenvalues of $C^{\mathsf T}C$ are $|f(k)|^2$, so its polar factor retains only the phase
\begin{equation}
q(k)=\frac{f(k)}{|f(k)|}.
\label{aeq:majorana-q}
\end{equation}

This can also be seen directly from the two-component momentum-space Hamiltonian. With
\begin{equation}
a_j=L^{-1/2}\sum_k e^{ikj}a_k,
\qquad
b_j=L^{-1/2}\sum_k e^{ikj}b_k,
\end{equation}
and $a_k^\dagger=a_{-k}$, $b_k^\dagger=b_{-k}$, one obtains
\begin{equation}
\mathcal H
=
\frac12\sum_k
\Psi_k^\dagger h(k)\Psi_k,
\qquad
\Psi_k=
\begin{pmatrix}
a_k\\
b_k
\end{pmatrix},
\qquad
h(k)=
\begin{pmatrix}
0&i f(k)^*\\
-i f(k)&0
\end{pmatrix}.
\end{equation}
The factor $1/2$ avoids double counting the Majorana bilinear. Since
\begin{equation}
h(k)^2=|f(k)|^2 I_2,
\end{equation}
the single-particle energies are $\pm|f(k)|$. A normalized negative-energy eigenvector and its projector are
\begin{equation}
u_-(k)
=
\frac{1}{\sqrt2}
\begin{pmatrix}
-iq(k)^*\\
1
\end{pmatrix},
\qquad
P_-(k)
=
\frac12
\begin{pmatrix}
1&-iq(k)^*\\
iq(k)&1
\end{pmatrix}.
\end{equation}
The ground-state projector therefore depends only on the phase $q(k)$; $|f(k)|$ sets the excitation energy but does not affect the occupied single-particle subspace. Eigenvector phase conventions drop out of the projector.

Fourier transforming the polar factor gives
\begin{equation}
G_{j\ell}
=
\frac1L\sum_k
q(k)e^{ik(j-\ell)},
\end{equation}
and translation invariance reduces it to the single function
\begin{equation}
g(r)
\equiv
\langle i b_j a_{j+r}\rangle
=
\frac1L\sum_k
\frac{f(k)}{|f(k)|}e^{-ikr}.
\label{aeq:majorana-gr}
\end{equation}
Indices crossing the boundary are understood using the antiperiodic extension, so
\begin{equation}
g(r+L)=-g(r).
\end{equation}
The $L$ values $g(0),\ldots,g(L-1)$ therefore determine every entry of $G$. Pairing the $k$ and $-k$ contributions shows that $g(r)$ is real.

In the original spin variables,
\begin{equation}
g(0)=-\langle\sx_j\rangle,
\end{equation}
while, for a string that does not cross the boundary,
\begin{equation}
g(r)
=
\left\langle
\sz_j
\sx_{j+1}\cdots\sx_{j+r-1}
\sz_{j+r}
\right\rangle,
\qquad r\geq1.
\end{equation}
The autoencoder input therefore contains correlations over all separations, including the cluster-string correlations associated with the different phases, rather than a manually selected local order parameter.

\subsection{Why $\boldsymbol g$ completely specifies the Gaussian ground state}

Collect the Majoranas into
\begin{equation}
\boldsymbol\gamma
=
(b_1,\ldots,b_L,a_1,\ldots,a_L)^{\mathsf T}
\end{equation}
and define the covariance matrix
\begin{equation}
\Gamma_{mn}
=
\frac{i}{2}
\langle[\gamma_m,\gamma_n]\rangle.
\end{equation}
For the ground state above,
\begin{equation}
\Gamma=
\begin{pmatrix}
0&G\\
-G^{\mathsf T}&0
\end{pmatrix},
\qquad
\langle\gamma_m\gamma_n\rangle
=
\delta_{mn}-i\Gamma_{mn},
\qquad
\Gamma^2=-I_{2L}.
\label{aeq:majorana-cov}
\end{equation}
The same-species correlations are
\begin{equation}
\langle a_j a_\ell\rangle
=
\langle b_j b_\ell\rangle
=
\delta_{j\ell},
\end{equation}
so all nontrivial two-point information is contained in $G$.

A ground state of a nonsingular quadratic Hamiltonian is a fermionic Gaussian state. Wick's theorem therefore expresses every higher even-order correlation function in terms of the two-point correlations, while odd-order correlations vanish in a state of definite fermion parity. Consequently, $\Gamma$ determines the ground-state density operator. Translation invariance and the bipartite form of the Majorana couplings further reduce $\Gamma$ to the $L$ real numbers
\begin{equation}
\boldsymbol g=(g(0),\ldots,g(L-1)).
\end{equation}
Thus no ground-state information is discarded by using $\boldsymbol g$ instead of the many-body wavefunction within this Gaussian family and the chosen boundary sector.

There is also a natural normalization. Discrete Fourier orthogonality and $|q(k)|=1$ give
\begin{equation}
\|\boldsymbol g\|_2^2
=
\sum_{r=0}^{L-1}|g(r)|^2
=
\frac1L\sum_k|q(k)|^2
=
1.
\label{aeq:majorana-norm}
\end{equation}
The autoencoder therefore receives a real unit vector of length $L$.

\subsection{Implementation and use in the scaling benchmark}

For the scaling benchmark we take $L=500$ and vary the coupling dimension up to $d=200$. Each ground-state sample nevertheless contains only the $L=500$ components of $\boldsymbol g$. A local training cloud consists of $30$ such vectors, and the autoencoder widths are $500$--$32$--$16$--$32$--$500$ (Table~\ref{tab:hyper}).

A sample is constructed directly by evaluating $f(k_n)$ at the $L$ antiperiodic momenta, forming $q(k_n)=f(k_n)/|f(k_n)|$, and applying the inverse discrete Fourier transform in Eq.~\eqref{aeq:majorana-gr}. No many-body wavefunction or $2^L$-dimensional Hilbert-space representation is constructed, and no numerical ground-state diagonalization is required. At $L=500$ this evaluation takes milliseconds in our implementation. As a numerical check, we compared Eq.~\eqref{aeq:majorana-gr} with an explicit polar decomposition of $C$ and found agreement to $2\times10^{-14}$.

The operations used to construct $\boldsymbol g$ are differentiable with respect to the couplings wherever $f(k_n)\neq0$. This permits the ground-state representation to be used directly within the autonomous trajectory search. The overlap-based reconstruction loss applied to these normalized inputs should therefore be interpreted as a similarity measure between ground-state correlation vectors, not as the fidelity between many-body wavefunctions.

It is useful to distinguish the two roles played by $f(k)$ in the benchmark. Its normalized phase $q(k)=f(k)/|f(k)|$ determines the ground-state correlation vector $\boldsymbol g$ supplied to the encoder, while its winding number provides the exact topological phase label used only for evaluating phase coverage. Although both quantities originate from the same function $f(k)$, the winding label is never supplied to the encoder or used as a preselected phase classifier. The autonomous system operates directly on the ground-state correlations.

Finally, the reduction of the ground state to $\boldsymbol g$ relies on the quadratic Gaussian structure, translation invariance, and the bipartite Majorana coupling of this benchmark. It should not be interpreted as a complete state representation for a generic interacting chain.

\section{Structure and complexity of the Chern inverse problem}
\label{app:inverse}

Section~\ref{subsec:results_chern} uses the generalized QWZ model as a benchmark in which the accessible phases are not known in advance. Here we make this statement more precise. The forward problem---evaluating the Chern number $\mathcal C(\boldsymbol{\theta})$ for a given parameter point---admits a simple signed-count formulation, whereas no general inverse construction is known that maps a prescribed Chern number $\mathcal C^\star$ to a parameter point $\boldsymbol{\theta}$ realizing it. We derive a general upper bound on $|\mathcal C|$, examine simple harmonic constructions, and discuss the structure and complexity of the inverse problem.

\subsection{Reduction to a signed lattice count}

In Eq.~\eqref{eq:chernmodel}, $d_x$ depends only on $k_x$ and $d_y$ only on $k_y$. The joint zero set $\{d_x=d_y=0\}$ is therefore the product of the two root sets, and the Jacobian of $(d_x,d_y)$ is diagonal there. Taking the north pole of $S^2$ as a regular value, the degree is the signed count of its preimages,
\begin{equation}
\mathcal C
=
-\!\!\sum_{(i,j):\,d_z(k_x^i,k_y^j)>0}\!\!
\operatorname{sign}d_x'(k_x^i)\,
\operatorname{sign}d_y'(k_y^j),
\label{aeq:degree}
\end{equation}
where $\{k_x^i\}$ and $\{k_y^j\}$ are the zeros of $d_x$ and $d_y$ in $[0,2\pi)$. We verified Eq.~\eqref{aeq:degree} against the lattice evaluation used in Sec.~\ref{subsec:results_chern} for random parameter points at $H=16$, finding exact agreement for all points tested.

Two useful consequences follow. For simple zeros, which are generic, the derivative of $d_x$ alternates in sign between consecutive zeros, and likewise for $d_y$. The factor
\begin{equation}
\operatorname{sign}d_x'(k_x^i)\,
\operatorname{sign}d_y'(k_y^j)
\end{equation}
therefore forms a checkerboard pattern on the product grid. The Chern number is the corresponding signed checkerboard sum over the subset of grid points satisfying $d_z>0$.

Moreover, a trigonometric polynomial $\sum_{n=1}^{H}a_n\sin(nk)$ has at most $2H$ zeros in $[0,2\pi)$, so the product grid contains at most $4H^2$ points. Because the checkerboard contains equal numbers of positive and negative contributions, the degree obeys
\begin{equation}
|\mathcal C|\leq 2H^2.
\label{aeq:chern-bound}
\end{equation}
For $H=16$, this gives
\begin{equation}
|\mathcal C|\leq512,
\end{equation}
corresponding to $1025$ integer values between $-512$ and $512$. This is only an upper bound: it does not imply that every value in this interval, or the extremal values themselves, can actually be realized by Eq.~\eqref{eq:chernmodel}.

\subsection{Sign patterns at fixed harmonics}

Once the zeros of $d_x$ and $d_y$ are fixed, the remaining parameters enter Eq.~\eqref{aeq:degree} only through the signs of $d_z$ on this grid. At each grid point,
\begin{equation}
d_z
=
m+\sum_{n=1}^{H}c_n
\bigl(\cos(nk_x)+\cos(nk_y)\bigr)
+e\cos(k_x)\cos(k_y)
\end{equation}
is linear in the $(H+2)$-component parameter vector
\begin{equation}
(c_1,\ldots,c_H,m,e).
\end{equation}
For the generic grids considered here, we confirmed that the corresponding design matrix has full rank $H+2$.

Each grid point therefore defines a hyperplane through the origin of $\mathbb R^{H+2}$, and the achievable sign patterns correspond to the full-dimensional cells of this hyperplane arrangement. Because the hyperplanes are central, $N$ such hyperplanes in $\mathbb R^D$ divide the space into at most
\begin{equation}
2\sum_{k=0}^{D-1}\binom{N-1}{k}
\end{equation}
cells~\cite{buck1943}. At $H=16$, a maximal $32\times32$ zero grid has $N=1024$ points and $D=18$, reducing the naive $2^{1024}\approx10^{308}$ possible sign assignments to at most approximately $2\times10^{38}$ cells.

Thus, at fixed $d_x$ and $d_y$, not every assignment of signs to the zero grid is realizable. Membership of a particular sign pattern can nevertheless be checked by a linear feasibility problem. As a numerical illustration, fixing one generic choice of $(a_n,b_n)$ and sampling only the $18$ parameters entering $d_z$ yields $110$ distinct Chern numbers spanning $[-69,58]$ after $2\times10^5$ samples. The number of observed values continues to grow slowly with additional samples, illustrating that even a fixed zero grid can support a nontrivial set of sectors.

\subsection{Simple harmonic constructions and their limitations}

A particularly transparent family is obtained by retaining a single harmonic $n$ and setting
\begin{equation}
a_n=b_n=c_n=1,
\end{equation}
with all other harmonic coefficients set to zero and the mass $m$ varied. In this case the momentum-space texture consists of an $n\times n$ repetition of the basic QWZ texture, and the resulting phases include
\begin{equation}
\mathcal C=\pm n^2.
\end{equation}
For $H=16$, this construction therefore reaches Chern numbers with magnitude as large as
\begin{equation}
H^2=256,
\end{equation}
while the general bound in Eq.~\eqref{aeq:chern-bound} permits $|\mathcal C|$ as large as $512$. Whether the latter bound can be saturated within the full family is not known.

Adding harmonics enlarges the set of accessible Chern numbers, but there is no simple composition rule. In particular, the Chern numbers associated with individual harmonics do not simply add when those harmonics are combined. For example, combining harmonics $2$ and $3$ in the simple unit-coefficient ansatz yields $\mathcal C=8$ rather than $2^2+3^2=13$, while combining harmonics $1$ and $16$ yields $\mathcal C=44$ rather than $1^2+16^2=257$. Thus knowledge of the sectors realized by individual harmonics does not provide an inverse construction for their combinations.

Enumerating the unit-coefficient ansätze $a_n=b_n=c_n=1$ with one, two, and three active harmonics and sweeping the mass realizes $33$, $99$, and $174$ distinct Chern numbers, respectively. Adding a fourth harmonic produces little further increase. These values remain far below the $1025$ integer Chern numbers permitted by the general bound $|\mathcal C|\leq512$.

The restricted coverage of these simple constructions is strongly influenced by symmetry. Setting $a_n=b_n$ makes $d_x$ and $d_y$ identical functions of their respective momenta, so their root sets coincide and the grid in Eq.~\eqref{aeq:degree} is symmetric under interchange of its two axes. The mass function $d_z$ is likewise symmetric under $k_x\leftrightarrow k_y$ and even under $\mathbf k\rightarrow-\mathbf k$. These constraints restrict the sign patterns that can be realized on the checkerboard grid and consequently reduce the accessible Chern numbers.

Breaking this symmetry increases the range of sectors reached. At a fixed number of evaluations, replacing the unit coefficients by generic coefficients on the same active harmonics separates the $d_x$ and $d_y$ root sets. For two active harmonics, the number of observed Chern values increases from $88$ spanning $[-176,160]$ to $123$ spanning $[-256,256]$; for three active harmonics, it increases from $159$ spanning $[-159,153]$ to $195$. Generic coefficients therefore reach substantially more sectors than the symmetric closed-form families.

This increased coverage, however, does not provide an inverse construction. The generic coefficients themselves must be searched, and the resulting Chern number still has to be evaluated afterward. We do not know a general rule that, given a target $\mathcal C^\star$, directly produces a coupling vector $\boldsymbol{\theta}$ satisfying $\mathcal C(\boldsymbol{\theta})=\mathcal C^\star$.

\subsection{Relation to the sectors found by autonomous search}

The simple constructions above provide useful reference points for assessing how much of the phase space can be reached by readily specified parameter families, but they are not used as a baseline for the scaling comparison in Sec.~\ref{subsec:results_chern}. That comparison is exclusively between the autonomous search and random sampling at matched query budgets.

Pooling the five autonomous-search runs reported in Sec.~\ref{subsec:results_chern}, the search reaches $283$ distinct Chern sectors spanning
\begin{equation}
\mathcal C\in[-192,188].
\end{equation}
Of these, $151$ are also realized by the one-, two-, and three-harmonic constructions considered above, whereas $132$ are not reached by those constructions. The latter should not be interpreted as Chern sectors for which no realization exists---the autonomous search itself provides explicit realizations---but rather as sectors not captured by these simple constructive families.

The largest magnitude reached by the autonomous search, $|\mathcal C|=192$, lies within the range already accessible to simple constructions. The role of the search is therefore not to extend the known extremal Chern number in this family, but to efficiently locate a much broader collection of sectors throughout the accessible range. In particular, it supplies explicit coupling vectors for many Chern values that are not produced by the simple harmonic ansätze considered here.

\subsection{Complexity and open questions}

The level set
\begin{equation}
\{\boldsymbol{\theta}:
\mathcal C(\boldsymbol{\theta})=\mathcal C^\star\}
\end{equation}
can be represented through polynomial sign conditions in the $3H+2$ couplings together with auxiliary variables specifying the roots of the trigonometric polynomials. The corresponding realizability question can therefore be formulated within the existential theory of the reals and is decidable in $\mathrm{PSPACE}$~\cite{canny1988,bpr2006}. This provides an upper bound on the decision complexity; it does not establish that the inverse problem is $\exists\mathbb R$-hard or $\mathrm{PSPACE}$-hard.

A complete characterization of the realizable Chern numbers remains unavailable. The positions of the checkerboard grid are themselves constrained: they arise as zeros of trigonometric polynomials of degree $H$ and vary continuously with $(a_n,b_n)$. Consequently, the grid geometry and the sign pattern imposed by $d_z$ cannot be chosen independently. We know of no general characterization of which combinations are simultaneously realizable and no general construction that maps a prescribed Chern number to a corresponding parameter vector.

These observations clarify the sense in which the inverse problem is open in the benchmark. The forward map $\boldsymbol{\theta}\mapsto\mathcal C$ is efficiently evaluable, and several restricted families admit simple constructions, but the complete set of accessible Chern sectors and a general inverse map $\mathcal C^\star\mapsto\boldsymbol{\theta}$ are not known. It is this gap---rather than the computational cost of evaluating $\mathcal C$---that motivates the generalized QWZ model as an open-ended phase-discovery benchmark.

\end{document}